\documentstyle[preprint,aps,epsf]{revtex}

\def\ket#1{\left | {#1} \right >}
\def\bra#1{\left < {#1} \right |}

\def\ie{{ i.e.}}

\def\pr{Phys. Rev.}
\def\prb{Phys. Rev. B }
\def\prl{Phys. Rev. Lett. }

\def\mplb{Mod. Phys. Lett. B}
\def\beq{\begin{equation}}
\def\eeq{\end{equation}}

\begin{document}

\draft
\title{Wigner delay time from a random passive and active medium}
\author{Sandeep K. Joshi$^*$\cite{jos}, Abhijit Kar Gupta$^{\dag}$\cite{abhie} and  A. M. Jayannavar$^*$\cite{amje}}
\address{$^*$Institute of Physics, Sachivalaya Marg, Bhubaneswar 751 005, India}
\address{$^{\dag}$Institute of Mathematical Sciences, Taramani, Chennai 600 113, India}

\maketitle

\begin{abstract}

We consider the scattering of electron by a one-dimensional random
potential (both passive and active medium) and numerically obtain the
probability distribution of Wigner delay time ($\tau$). We show that in a
passive medium our probability distribution agrees with the earlier
analytical results based on random phase approximation. We have extended
our study to the strong disorder limit, where random phase approximation
breaks down.  The delay time distribution exhibits the long time tail
($1/\tau^2$) due to resonant states, which is independent of the nature of
disorder indicating the universality of the tail of the delay time
distribution. In the presence of coherent absorption (active medium) we show that
the long time tail is suppressed exponentially due to the fact that the
particles whose trajectories traverse long distances in the medium are
absorbed and are unlikely to be reflected. 

\end{abstract}

The concept of delay time which was introduced by Wigner\cite{wigner} way back in 1955
has received renewed interest in recent years\cite{jayan,pendry,hein,yan,gopar,comtet,joshi1}. The delay time in the
scattering process is generally taken to be related to the duration of
collision event or the time spent by the particle (the wave packet) in the
region of interaction. Very recently universal parametric correlations of
phase shifts and delay times in quantum chaotic scattering regimes have
been reported\cite{yan}. By assuming the validity of random matrices for describing
the statistics of closed chaotic systems the distribution of inverse delay
time is given by the Laguerre ensemble from random matrix theory. The
delay time statistics is intimately connected with the issue of dynamic
admittance of microstructures\cite{thomas}. The case study being the mesoscopic
capacitance\cite{nku} fluctuation for the case of a chaotic cavity coupled
capacitively to the backgate\cite{gopar}. This has involved calculation of
delay time given by the energy derivative of the phase associated with the
scattering matrix.

It may be noted that for the case of a single channel the distribution of
the delay time for a disordered semi-infinite sample has been obtained
earlier by using the invariant imbedding approach\cite{jayan}. The stationary
distribution $P_s(\tau)$ for dimensionless delay time $\tau$ is given by

\beq
P_s(\tau)~=~\frac{\lambda e^{\lambda
tan^{-1}\tau}}{(e^{\lambda\pi/2}-1)(1+\tau^2)},
\label{pst}
\eeq

\noindent where $\lambda$ is proportional to the disorder induced
localization length and the most probable value of $\tau$ occurs at
$\tau_{max}=\lambda/2$. At long time tail of the above distribution scales
as $1/\tau^2$. The average value of $\tau$ is
logarithmically divergent indicating the possibility of the particle
traversing the infinite sample before being totally reflected, presumably
due to the Azbel resonances\cite{azbel}, which makes Landauer's four probe conductance
infinite even for a finite sample. If the disordered region is
semi-infinite, the reflection coefficient will be unity, and the complex
reflection amplitude will have the form $R~=~e^{i\theta(E)}$. If the wave
packet is incident on the disordered sample it will not be immediately
reflected back into the lead region, but will be delayed by time
proportional to $\tau~=~\hbar d\theta/dE$. This energy dispersive
backscattering (or energy dependent random time delay) leads to
non-cancellation of the instantaneous currents at the surface involving
the incident and reflected waves. This non-cancellation is expected to
lead to a low temperature $1/f$ type noise that should be universal
\cite{jayan,pendry,hein}. A very recent
study claims the delay time distribution in one-channel case to be
universal and is independent of the nature of disorder (especially the
long time tail distributions)\cite{comtet}. We would like to examine this claim through
our study. We would like to emphasize that to obtain $P_s(\tau)$ (Eqn.
\ref{pst}) earlier studies invoke several approximations such as random
phase approximation (RPA), which is only valid in the small disorder
regime and moreover, correlation between the phase and the delay time is
neglected. 

In view of the recent claim of universality in the delay time
distribution, we have studied numerically the one-dimensional disordered
medium and computed the $P_s(\tau)$. Our results agree qualitatively with
those obtained earlier even in the strong disorder limit where random
phase approximation breaks down. We also observe same scaling behavior of
the tail of the stationary delay time distribution for different
distribution of the disorder potential. We have also studied the delay
time distribution in a random active medium, \ie, in the presence of coherent
absorption. In this case we observe that $1/\tau^2$ long time tail distribution
is strongly suppressed. This is due to the fact that in the absence of
absorption the tail distribution arises from the possibility of particle
traversing the infinite sample (due to resonances) before being reflected\cite{jayan,azbel}. 
As these paths traverse a long distance, in the presence of absorption the
particles will get absorbed in the medium and they cannot be reflected. 
Consequently distribution at the tail gets exponentially suppressed. There
are several physical situations where one encounters the absorption of
elementary particles (electrons, excitons or magnetic excitations, etc.)
due to impurities or trapping centers in a medium. One recent example
being the light (photon) propagation in a lossy dielectric medium\cite{abhi,pradhan,amjabscoh}.  

To describe the motion of a quasi particle on a lattice, we use the 
following Hamiltonian in a tight-binding one-band model:

\beq
H~=~\sum~\epsilon_n\ket{n}\bra{n}~+~V(\ket{n}\bra{n+1}~+~\ket{n}\bra{n-1}),
\label{ham}
\eeq

\noindent where $\ket{n}$ is the non degenerate Wannier orbital at site $n$, 
$\epsilon_n$ is the site energy at the site $n$ and $V$ is the hopping 
matrix element connecting nearest neighbors separated by a unit lattice 
spacing. We consider three different kinds of disorder where the site 
energies $\epsilon_n$ are assumed uncorrelated random variables having 
distributions which are uniform ($P(\epsilon_n)=1/W$), Gaussian 
($P(\epsilon_n) \propto e^{-\epsilon_n^2/W^2}$) and exponential 
($P(\epsilon_n) \propto e^{-\epsilon_n/W}$). In case of modeling absorption we 
make the site energies complex ($\epsilon_n \equiv \epsilon_n + i\eta$)
with the imaginary part $\eta$ chosen to be spatially constant (coherent
absorption)\cite{abhi}. 
The $N$ site ($n~=~1~to~N$) disordered 1D sample is embedded in a 
perfect infinite lattice having all site energies zero. The 
well known transfer-matrix method\cite{abhi,trm,joshi} is used to calculate the reflection 
amplitude $r(E)~=~|r|e^{-i\theta(E)}$ and its phase $\theta(E)$ at 
two values of incident energy $E=E_0\pm\delta E$.
The delay time is then calculated using the definition 
$\tau = \hbar d\theta/dE$. Throughout our following discussion we consider 
delay time $\tau$ in a dimensionless form by multiplying it with $V$ and we 
set $\hbar=m=1$.

In view of the fact that the value of incident energy $E_0$ will not change the physics of
the problem, in the following we choose $E_0=0$ and $dE=2\delta E=0.002$.  In
calculating the distribution of various quantities we have taken at least
10 000 realizations of the disordered sample. The disorder
strength ($W$) and absorption strength ($\eta$) are scaled with respect to
$V$ i.e., $W \equiv W/V$ and $\eta \equiv \eta/V$. To calculate the 
stationary distribution of delay time we have considered a sample of 
length ($L$) 5 times the localization length ($\xi$), where the localization 
length is calculated by a standard prescription\cite{econ}. We have confirmed
that by changing the length $L$ of the sample the distribution remains unchanged. 

\begin{figure}[t]
\protect\centerline{\epsfxsize=4in \epsfbox{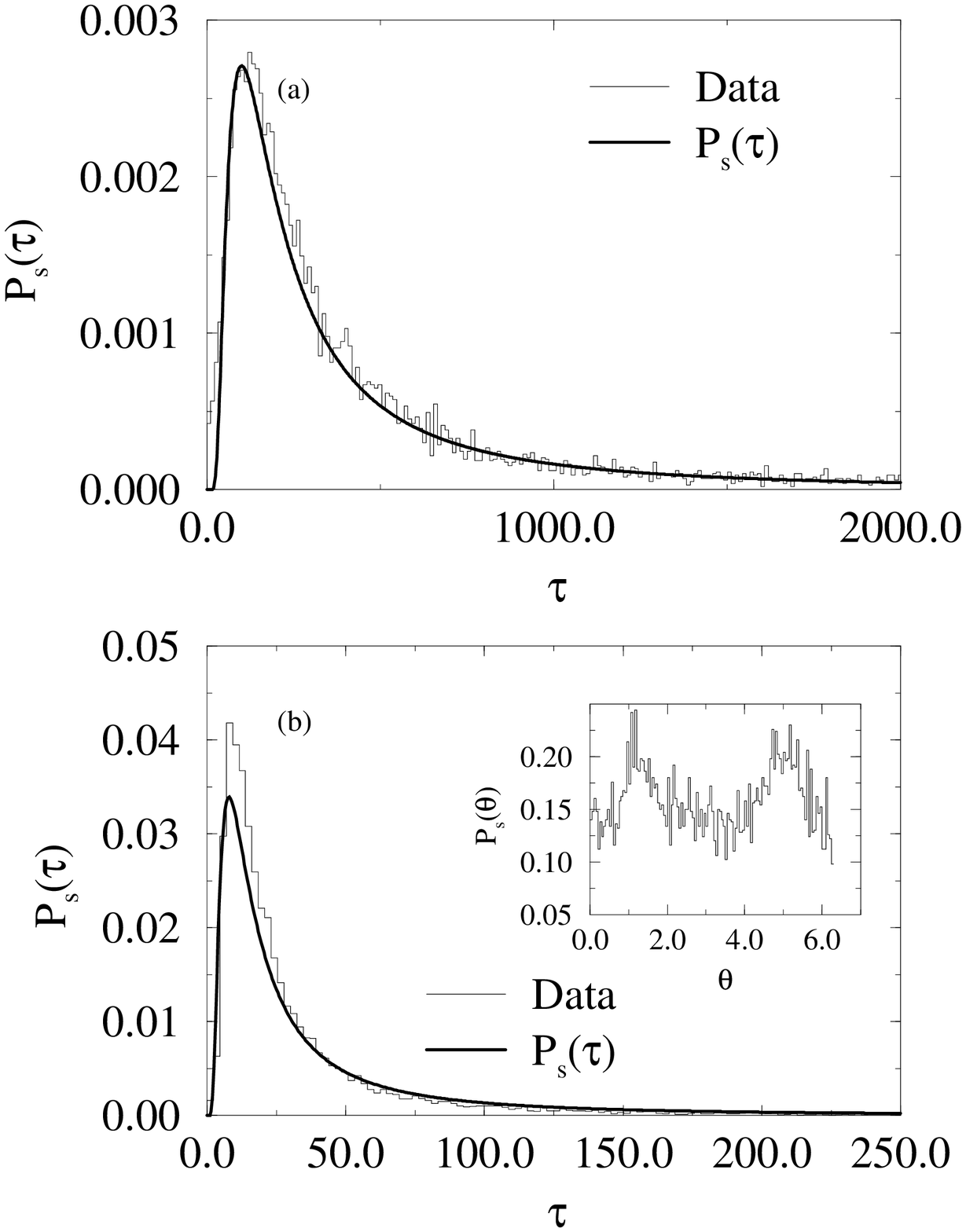}}
\caption{The stationary distribution of delay time $P_s(\tau)$ for
(a) weak disorder (W=0.5) and (b) strong disorder (W=2.0).}
\label{pstfig}
\end{figure}

\begin{figure}[t]
\protect\centerline{\epsfxsize=4in \epsfbox{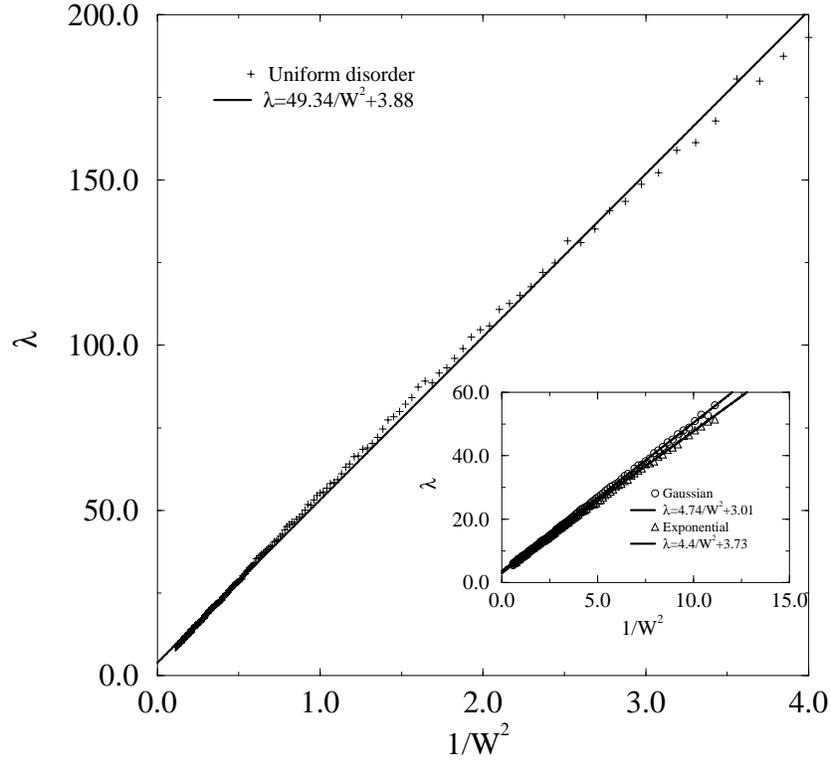}}
\caption{Plot of $\lambda$ versus $1/W^2$ for uniform disorder. The 
numerical fit is valid only for small values of disorder strength i.e.,  
$W \ll 1.0$. The inset shows the plot of $\lambda$ versus $1/W^2$ for 
Gaussian and exponential disorder.} 
\label{lvsw} 
\end{figure}

In Fig. \ref{pstfig} (a) and (b) we show the plot of numerical data (thin 
line) for the stationary distribution $P_s(\tau)$ of delay time $\tau$ 
for weak disorder ($W=0.5$) and strong disorder ($W=2.0$) respectively. 
The thick line in the figure is the numerical fit obtained by using the 
expression for $P_s(\tau)$ given in Eqn. \ref{pst}. We see that the fit 
is fairly good even for strong disorder ($W=2.0$) for which the stationary
distribution of the phase of reflected wave, $P_s(\theta)$, shows (inset of Fig. \ref{pstfig}
(b)) two distinct peaks 
indicating the failure of random phase approximation (RPA) in this 
regime\cite{stone,jayanpra}. The values of $\lambda$ thus obtained, when plotted against 
$1/W^2$, indicate that $\lambda$ scales as $1/W^2$ as shown in the Fig. 
\ref{lvsw}. Since the energy of incoming particles is fixed, the most 
probable traversal time will be proportional to the typical length 
traveled by the particle in the sample, \ie, to the localization length. 
As the localization length for passive disordered systems\cite{econ} scales as 
$1/W^2$ and $\lambda = \tau_{max}/2$, the proportionality of $\lambda$ to 
$1/W^2$ stands clear. In the inset of Fig. \ref{lvsw} we show the plot of
$\lambda$ versus $1/W^2$ for Gaussian and exponential disorder. We see 
that in the weak disorder limit $\lambda$ scales as $1/W^2$ for Gaussian and
exponential disorder as well.

\begin{figure}[t]
\protect\centerline{\epsfxsize=4in \epsfbox{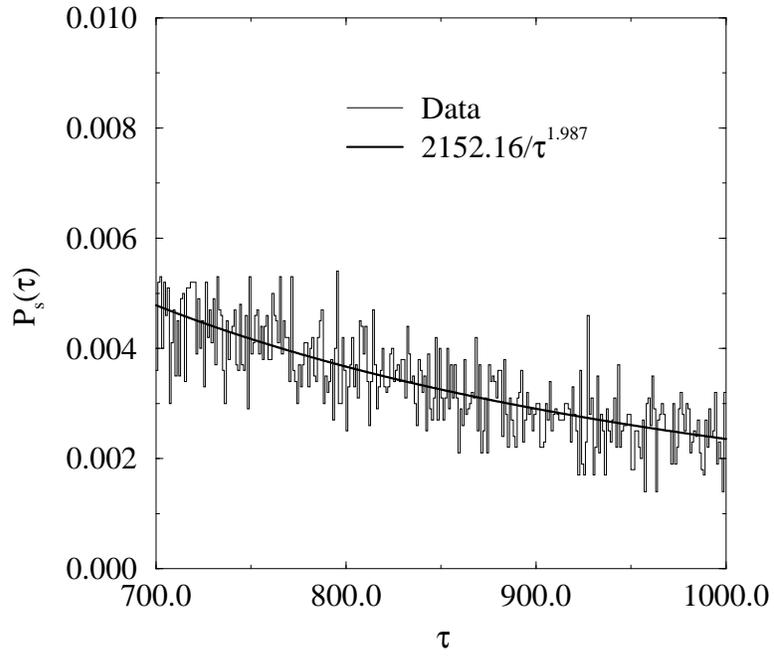}}
\caption{The plot of tail of $P_s(\tau)$ for $W=0.5$. Numerical fit 
(thick line) using expression $\alpha/\tau^\beta$ yields 
$\beta \approx 2.0$.}
\label{tdist} 
\end{figure}

We would now like to take a closer look at the tail of the stationary 
distribution $P_s(\tau)$. As mentioned earlier, within RPA the long time delay 
distribution would scale as $1/\tau^2$. The numerical least square 
fit of the expression $\alpha/\tau^\beta$ to the long time tail data 
gives $\beta \approx 2$. This is shown in Fig. 
\ref{tdist}. The appearance of such a tail is attributed to the presence 
of resonant states is due to certain realizations, as in these cases the particle
travels a long distance before getting reflected\cite{jayan}.

Motivated by the recent claim\cite{comtet} of universality of delay time distribution 
in one-channel case (within RPA), we numerically study the delay 
distribution $P_s(\tau)$ for the case of Gaussian and exponential 
disorder. The expression in Eqn. \ref{pst} was derived within RPA using a 
Gaussian, delta-correlated random potential with zero mean. We have 
already seen that it fits very well with the data for $P_s(\tau)$ in case 
of uniform disorder. Similarly, The agreement of Eqn. \ref{pst} with 
numerical data for Gaussian and exponential disorder is also excellent
in the weak disorder limit. 

\begin{figure}[t]
\protect\centerline{\epsfxsize=4in \epsfbox{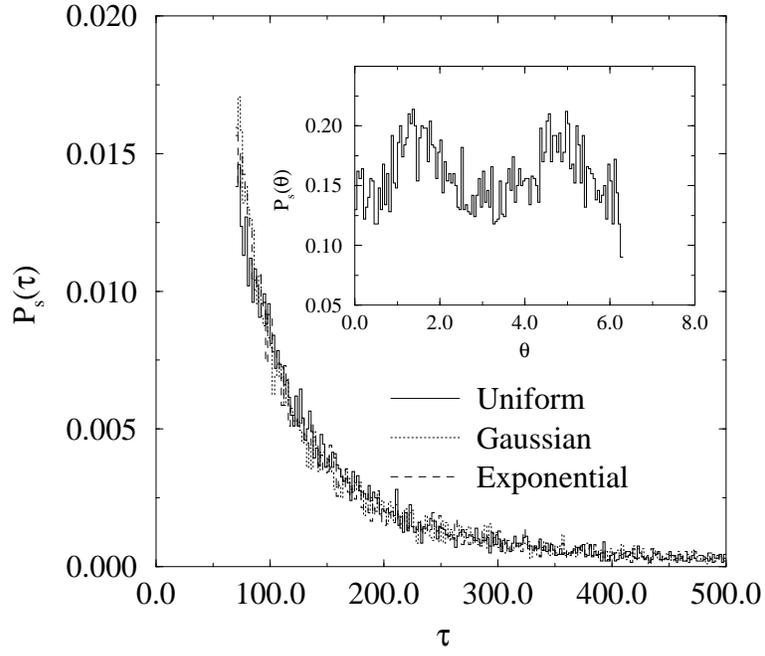}}
\caption{The plot of tail of $P_s(\tau)$ for the case of uniform, 
Gaussian and exponential disorder. The disorder strength in all three 
cases is $W=1.0$.} 
\label{univ} 
\end{figure}

We now look at the tail of delay distribution and its universality for the three different 
kinds of disorder beyond RPA. Since the origin of tail is due to the 
appearance of resonant realizations which are independent of strength and 
the type of disorder, we expect that the tail distribution would be universal 
beyond RPA too. In Fig. \ref{univ} we plot the tail distribution of $P_s(\tau)$
for uniform, Gaussian and exponential disorder characterized by the strength
$W=1.0$. It can be readily noticed that in all the cases rescaled graphs fall
on same curve (within our numerical error) clearly indicating the universal
nature of tail distribution.
For the value $W=1.0$, we are in a regime beyond RPA as can be seen from 
the non-uniformity of the stationary distribution $P_s(\theta)$ of the phase 
of the reflected wave shown in the 
inset of the Fig. \ref{univ}.Therefore, our numerical simulation 
results suggest the existence of universality in the long time tail 
distribution beyond RPA.

\begin{figure}[t]
\protect\centerline{\epsfxsize=4in \epsfbox{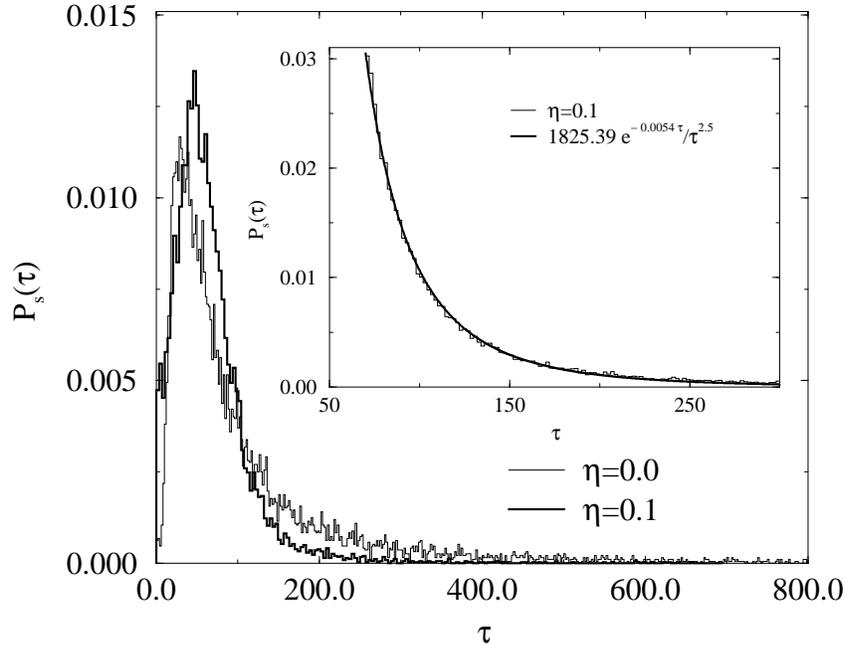}}
\caption{The plot of $P_s(\tau)$ for passive ($\eta=0$, shown with thin 
line) and active ($\eta=0.1$, shown with thick line) disordered 
medium. The inset shows the tail of $P_s(\tau)$, numerical data using 
thin line and numerical fit using thick line, for the absorbing medium.}
\label{active} 
\end{figure}

Finally we consider the stationary delay distribution for a coherently 
absorbing random medium. We take the absorption strength $\eta$ to be $0.1$
and the disorder strength $W$ to be $1.0$ for the active random medium.
From Fig. \ref{active} we see that the delay distribution for active 
(coherently absorbing) medium (shown with thick line) falls off much rapidly 
as compared to that of passive medium(shown with thin line). The numerical fit of the tail 
distribution for the active medium, shown in the inset of Fig. 
\ref{active}, indicates that the fall off is exponential as 
$Ae^{-\alpha \tau}/\tau^{2.5}$. This can be 
interpreted in terms of suppression of the resonant realizations due to  
the presence of absorption. In case of these resonant realizations, the 
time spent by the particle inside the sample is large as it travels large 
distance before getting reflected. This enhances the probability of the 
particle getting trapped or absorbed (exponentially) in the traversal length 
thereby suppressing the tail. Analytical study of the distribution of 
delay time in the presence of coherent absorption is under active 
consideration by us. 

In conclusion, we have numerically studied the universality of the 
long time tail of the stationary delay time distribution for a 
disordered one dimensional sample. Our study reveals that the 
universality (with respect to the type of disorder) holds for weak
as well as strong disorder, \ie, irrespective of the validity of the 
RPA. Also the typical long tail ($\sim 1/\tau^2$) in the stationary 
distribution gets exponentially suppressed if we consider a coherently 
absorbing random medium. 

\centerline{{\bf ACKNOWLEDGEMENTS}}

One of us (A.K.G) thanks Institute of Physics, Bhubaneswar, for hospitality,
where a major part of this work was carried out.

\end{document}